\documentclass[aps,prd,twocolumn,nofootinbib,preprintnumbers,superscriptaddress,preprintnumbers,
floatfix,10pt]{revtex4-2}

\usepackage{amsmath,amsfonts,amssymb,slashed,braket,bbold,physics}
\usepackage{cancel}

\usepackage{graphicx,epstopdf}

\usepackage{makecell}

\usepackage{xcolor}
\usepackage[normalem]{ulem} 

\definecolor{OliveGreen}{cmyk}{0.64,0,0.95,0.40}

\usepackage{silence}
\usepackage[colorlinks=true,
            linkcolor=blue,
            urlcolor=blue,
            citecolor=green,          
            bookmarks=true,
            bookmarksnumbered=true,
            breaklinks=true,
            pdfpagemode=Fullscreen,
            pdfstartview=FitBH]{hyperref}
\usepackage{orcidlink}

\definecolor{lime}{HTML}{A6CE39}
\DeclareRobustCommand{\orcidicon}{%
	\begin{tikzpicture}
	\draw[lime, fill=lime] (0,0) 
	circle [radius=0.16] 
	node[white] {{\fontfamily{qag}\selectfont \tiny ID}};
	\draw[white, fill=white] (-0.0625,0.095) 
	circle [radius=0.007];
	\end{tikzpicture}
	\hspace{-2mm}
}

\usepackage{placeins} 

\definecolor{gesfblack}{rgb}{0,0,0}

\definecolor{gesfblue}{rgb}{0.08,0.42,0.76}

\definecolor{gesfgreen}{rgb}{0,1,0}

\definecolor{gesfgrey}{rgb}{0.5,0.5,0.5}

\definecolor{gesflanse}{rgb}{0.00,0.50,0.50}

\definecolor{gesfpurple}{rgb}{0.47,0.19,0.42}

\definecolor{gesfred}{rgb}{1,0,0}

\definecolor{gesfwhite}{rgb}{1,1,1}

\definecolor{gesfyellow}{rgb}{0.7,0.4,0.3}

\newcommand{\gsec}[1]{{\hypersetup{linkcolor=red}Sec.\,\ref{#1}\hypersetup{linkcolor=blue}}}

\newcommand{\geqn}[1]{\hypersetup{linkcolor=blue}Eq.\,(\ref{#1})\hypersetup{linkcolor=blue}}
\newcommand{\gfig}[1]{{\hypersetup{linkcolor=violet}Fig.\,\ref{#1}\hypersetup{linkcolor=blue}}}
\newcommand{\gtab}[1]{{\hypersetup{linkcolor=gesflanse}Table\,\,\ref{#1}\hypersetup{linkcolor=blue}}}

\graphicspath{{figs/}}

\begin{document}

\title{RG Running of Multiple Neutrino Mixing Parameters at Oscillation Experiments}

\author{Peter B.~Denton}
\email{pdenton@bnl.gov}
\thanks{\href{https://orcid.org/0000-0002-5209-872X}{\orcidicon}}
\affiliation{High Energy Theory Group, Physics Department \\ Brookhaven National Laboratory, Upton, NY 11973, USA}

\author{Shao-Feng Ge}
\email{gesf@sjtu.edu.cn}
\thanks{\href{https://orcid.org/0000-0003-3711-125X}{\orcidicon}}
\affiliation{State Key Laboratory of Dark Matter Physics, Tsung-Dao Lee Institute \& School of Physics and Astronomy, Shanghai Jiao Tong University, Shanghai 200240, China}
\affiliation{Key Laboratory for Particle Astrophysics and Cosmology (MOE) \& Shanghai Key Laboratory for Particle Physics and Cosmology, Shanghai Jiao Tong University, Shanghai 200240, China}

\author{Chui-Fan Kong}
\email{kongcf@ibs.re.kr}
\thanks{\href{https://orcid.org/0009-0007-7010-5085}{\orcidicon}}
\affiliation{Particle Theory and Cosmology Group (PTC), Center for Theoretical Physics of the Universe (CTPU),
Institute for Basic Science, Daejeon 34126, Republic of Korea}

\author{Pedro Pasquini}
\email{pasquini@ifi.unicamp.br}
\thanks{\href{https://orcid.org/0000-0002-1689-442X}{\orcidicon}}
\affiliation{Instituto de F\'isica Gleb Wataghin - Universidade Estadual de Campinas (UNICAMP), {13083-859}, Campinas SP, Brazil}

\begin{abstract}
If the new physics scale is within the energy scale of neutrino
oscillation experiments, it may lead to a renormalization group
(RG) running effect between
the production and detection processes as well as between
different experiments. It is then possible to use multiple 
neutrino oscillation experiments to disentangle the multiple
RG running parameters.
We investigate this effect in a general model-independent
sense for a variety of flavor structures in the context
of upcoming experiments DUNE-ND, JUNO-TAO, and FASER$\nu$2
that span a large range in neutrino energies and many
different flavor combinations.
We find strong sensitivity to the running effects of new physics
with combination of these experiments, especially the
possibility of addressing the non-trivial degeneracies.
\end{abstract}

\preprint{CTPU-PTC-26-17}

\maketitle 

\section{Introduction}
The neutrino sector is one of the most promising places
to look for new physics beyond the Standard Model of
particle physics \cite{Arguelles:2022tki}.
While considerable progress has already been made to measure the oscillation parameters and future progress is expected to measure all six parameters \cite{Denton:2022een,Denton:2025jkt},
the nature and source of neutrino masses still remain unresolved.
From the model building point of view, possible new physics
may appear at a wide variety of energy scales. It is often simplest to consider the new physics scale to be much higher
than all scales in an experiment, typically the weak scale,
in an effective field theory context.
However, this may miss important physics, especially if
the new physics happens at a lower scale.
For example, in a compelling new physics model with non-standard
interactions (NSI) \cite{Wolfenstein:1977ue,Proceedings:2019qno},
the new physics scale could be heavy $>100$\,GeV and it could
also be much lighter, lying within the energy scale of neutrino
experiments
\cite{Farzan:2015hkd,Farzan:2016wym,Babu:2017olk,Farzan:2017xzy,Denton:2018xmq,Bernal:2022qba,Abbaslu:2024hep}.

If there is new physics that arises at intermediate scales,
its effect will appear as not just effective operators
but also renormalization group (RG) running of
the effective parameters.
Such an effect can significantly
change the neutrino oscillation phenomena
\cite{Babu:2021cxe, Babu:2022non}.
The oscillation amplitude,
\begin{align}
  \mathcal{A}_{\beta \alpha}
\equiv 
  \sum_i U_{\beta i}(Q^2_d)
  e^{-iLm_i^2/2 E_\nu} U^*_{\alpha i}(Q^2_p),
  \label{eq:amplitude}
\end{align}
contains three parts that corresponds to the neutrino
production ($U^*_{\alpha i}$), propagation
($e^{-iLm_i^2/2 E_\nu}$), and detection
($U_{\beta i}$) processes. Conventionally, the
mixing matrix $U^*_{\alpha i}$ at production
is treated as the same as its counterpart
$U_{\beta i}$ at detection. However, this needs
not to be true if new physics appears at a scale
below the characteristic scale $Q^2_p$ for production
or $Q^2_d$ for detection. The two mixing matrices
$U(Q^2_p)$ and $U(Q^2_d)$ could be different from
each other in the presence of RG running
\cite{Ge:2023azz, Ge:2024ibn}.

For example, the neutrinos at accelerator experiments
are produced dominantly from $\pi^\pm$ or $\mu^\pm$
decays whose
momentum transfer $\sqrt{Q^2_p}$ is around 100\,MeV
but are detected by nuclei scatterings with momentum
transfer $\sqrt{Q^2_d}$ up to several GeV.
This then leads to a variation
of the effective neutrino mixing matrix between the
production and detection
\cite{Casas:1999tg,Antusch:2003kp,Antusch:2005gp,Xing:2006sp,Ray:2010rz,Luo:2012ce,Ohlsson:2012pg,Ohlsson:2013xva,Xing:2017mkx,Huang:2018wqp,Babu:2021cxe,Babu:2022non,Ge:2023azz,Ge:2024ibn,Mir:2025fae}.
Such momentum transfer mismatch can affect
the oscillation probabilities.
Especially, the RG running effect on the Dirac CP
phase can be tested at not just the long-baseline
accelerator experiments such as DUNE \cite{Ge:2023azz},
but also the short-baseline reactor experiment
such as JUNO-TAO \cite{Ge:2024ibn}.
In this paper, we focus on the expected future measurements
from the near detector (DUNE-ND) at DUNE \cite{DUNE:2021tad},
JUNO-TAO the short-baseline detector apart of JUNO
\cite{JUNO:2020ijm}, and FASER$\nu$2 at the LHC
\cite{Anchordoqui:2021ghd}. Comparing with
long-baseline experiments, the near detectors with
short baseline can uniquely probe the RG running
parameters without complication from the oscillation
phases.

In a concrete model, it is unlikely that the running affects only a single oscillation parameter.
A more natural case is that multiple parameters
can experience RG running simultaneously.
In this paper, we consider not just the Dirac
CP phase but also the three mixing angles.
Especially, we show the correlation
among the RG running parameters and possible
ways of breaking the degeneracy with multiple
experiments. In \gsec{sec:RG review},
we first review the phenomenological consequences
of RG running on the neutrino oscillations.
We then discuss the important role of considering multiple experiments at multiple different energies with different flavor channels accessible to address degeneracies in \gsec{sec:degeneracies}.
We present our numerical results in \gsec{sec:sensitivities} and then discuss them and conclude in \gsec{sec:conclusions}.

\section{Neutrino Oscillation with Multiple RG Running Parameters}
\label{sec:RG review}

Quantum radiative corrections can typically
introduce energy dependence which can be parameterized as RG running of the neutrino mixing parameters. 
To be more concrete, the running of the neutrino 
oscillation parameters are described 
by the $\beta$-functions $\beta_X$ and the renormalization 
equation \cite{Antusch:2005gp, Ray:2010rz},
\begin{equation}
    \frac{dX}{d\ln |Q^2|}
=
    \beta_X,
\label{eq:beta-function}
\end{equation}
where $X$ denotes the neutrino mixing 
parameters including the three mixing angles $\theta_{ij}$
with $\beta_{ij}$ and the Dirac CP phase $\delta_D$
with $\beta_\delta$.
Without loss of generality, we take the Lorentz-invariant 
momentum transfer $|Q^2|$ as the renormalization scale, which is 
known as the Gell-Mann-Low scheme \cite{Gell-Mann:1954yli,
Wu:2013ei} and widely used in the literature 
\cite{Bustamante:2010bf,Babu:2021cxe,Babu:2022non,
Ge:2023azz,Ge:2024ibn}. 

The $\beta_X$ term controls the running 
of $X$ and is possibly a function of the mixing 
parameters themselves \cite{Antusch:2005gp, Ray:2010rz}.
Below the corresponding new physics scale $Q^2_0$, 
$\beta(Q^2 < Q^2_0) = 0$ and there is no 
running. The RG running effect appears above
$Q^2_0$ when $\beta_X$ becomes nonzero.
For concreteness, the value of $Q^2_0$ can
originate from a hidden light mediator mass. 
In this case we generally expect $Q^2_0 \gtrsim 1$\,MeV$^2$ 
to avoid cosmological bounds \cite{Venzor:2020ova}, which means that there is quite a bit of interesting parameter space probed by neutrino experiments which are the MeV, GeV, and TeV scales.

For $Q^2 \geq Q^2_0$, the RG 
running effect of mixing parameters is 
typically small \cite{Ge:2024ibn} 
and the induced variation of 
$\beta_X$ does not exceed 10\% as a rough 
estimation. Thus, $\beta_X$ can be regarded
as a constant and perturbative variable
($\beta_X\lesssim \mathcal{O}(10^{-1})$)
for typical oscillation experiments. In the small $\beta_X$ 
regime and $Q^2 \geq Q^2_0$, the problem 
becomes model-independent with the solution of 
\geqn{eq:beta-function} obtained by a linear 
approximation,
\begin{equation}
  X(Q^2)
\approx 
  X(Q^2_0)
+ \beta_X \ln 
  \left(\left|\frac{Q^2}{Q^2_0}\right|\right).
\label{eq:LinearSolution}
\end{equation}
While the exact values of the new 
physics scale $Q^2_0$ and $\beta_X$ depend
on the concrete model \cite{Ray:2010rz,
Ohlsson:2013xva,Babu:2021cxe} in the linear regime, 
$Q^2_0$ and $\beta_X$ can be treated as phenomenological
parameters to be constrained through experimental data with $\beta_X=0$ returning to the standard physics picture.

The oscillation probability from the $\alpha$
flavor to the $\beta$ flavor can be obtained as
$P_{\alpha \beta}(L) \equiv |\mathcal A_{\beta \alpha}|^2$.
Besides the neutrino mixing parameters with
the RG running effect, the experimental setup
characterized by the neutrino energy $E_\nu$ and
the baseline $L$ can also affect the oscillation
probability via the evolution phase
$e^{- i L m^2_i / 2 E_\nu}$
as shown in \geqn{eq:amplitude}.
A prominent
feature of quantum interference is that it is the
phase differences $e^{- i L \Delta m^2_{ij} / 2 E_\nu}$
with the mass squared difference
$\Delta m^2_{ij}\equiv m^2_i-m^2_j$ that
finally determine the interference probability.

However, it is useful to look to environments where
the phase accumulation does not affect the observables
as this removes much of the complications arising
from the standard three-flavor oscillation parameters
not all of which are yet determined.
Notably, the evolution phases can reduce to 1 in 
the zero-distance limit of $L\ll 2E_\nu/\Delta m^2_{ij}$
\cite{Babu:2021cxe,Ge:2023azz,Ge:2024ibn}.
This limit can be achieved in the typical
short-baseline neutrino experiments and at
the near detectors of long-baseline experiments.
The oscillation probability then reduces to,
\begin{equation}
  P_{\alpha\beta}(L = 0)
=
\left|
  \left[ U_{d} U_{p}^\dagger \right]_{\beta \alpha}
\right|^2.
\label{eq:prob}
\end{equation}

The mismatch in the momentum transfers of the
neutrino production and detection processes will
manifest itself as mismatch,
$\delta U \equiv U_d - U_p$, between the two
mixing matrices. So \geqn{eq:prob} reduces to,
\begin{align}
  P_{\alpha \beta}(L=0)
=
\left|
  \left[
    \mathbb 1
  + \delta U U_p^\dagger
  \right]_{\beta \alpha}
\right|^2,
\label{eq:Pab0}
\end{align}
where we have used the fact that $U_p$ is a
unitary matrix, $U_p U^\dagger_p = \mathbb 1$.
As the zero-distance oscillation probability reduces
back to the usual case, $P_{\alpha \beta} (L = 0) \rightarrow \delta_{\alpha \beta}$ as $\delta U\to0$, it would naively seem that any deviation could appear with both linear and quadratic terms.

Actually, we see that the linear term vanishes and only
the quadratic order can survive in the zero-distance limit.
For $\alpha\neq \beta$, the unit matrix $\mathbb 1$
in \geqn{eq:Pab0} does not contribute and the
transition probability,
\begin{align}
  P_{\alpha\beta}(L=0)
=
\left|
  \left[ \delta U U_p^\dagger \right]_{\beta \alpha}
\right|^2,
\end{align}
is at the quadratic order only.
For $\alpha= \beta$, the transition probability is
\begin{align}
  P_{\alpha\alpha}(L=0)
=
  1
+ 2 \Re[(\delta U U^\dagger_p)_{\alpha\alpha}]
+ |(\delta U U^\dagger_p)_{\alpha\alpha}|^2,
\end{align}
with terms containing both the linear
and quadratic orders of the deviations $\delta U$
induced by RG running. However, the unitarity condition,
$U_{d} U_{d}^\dagger = \mathbb 1$, implies that
$2{\rm Re}[(\delta 
U U_{p}^\dagger)_{\alpha \alpha}] = - 
(\delta U \delta U^\dagger)_{\alpha \alpha}$.
In other words, the linear term cancels and the correction is always quadratic.

Putting things
together, the transition probability in the
zero-distance limit takes the form as,
\begin{align}
  P_{\alpha\beta}(L=0)
=
\left[
  1
- (\delta U \delta U^\dagger)_{\alpha \alpha}
\right] \delta_{\alpha \beta}
+
  [|\delta U U^\dagger_p|_{\beta \alpha}]^2,
\end{align}
from which one can clearly see that the deviation
from the usual $\delta_{\alpha \beta}$ is always at
the quadratic order.

Consequently, in the linear regime of 
\geqn{eq:LinearSolution}, the zero-distance 
oscillation probability is quadratic in 
$\ln^2\left(|Q^2_d|/|Q^2_p|\right)$ and
its scale dependence can be generally parametrized
in terms of effective parameters
$\mathcal B_{\alpha \beta}$,
\begin{align}
  P_{\alpha\beta}(L=0)
\equiv 
  \delta_{\alpha \beta} 
+ \mathcal B_{\alpha \beta} 
  \ln^2 \left( \frac{|Q^2_d|}{|Q^2_p|} \right),
\label{eq:LinearApprox_P0}
\end{align}
with $\delta_{\alpha \beta}$ being the Kronecker delta.
The effective running parameters $\mathcal B_{\alpha \beta}$
for the oscillation probability $P_{\alpha \beta}(L=0)$
are functions of the mixing angles and CP phases 
\textit{calculated at production}. Note that the
running of $P_{\alpha \beta}(L=0)$ is governed by
a double logarithm rather than the single one in
\geqn{eq:LinearSolution}.

While $P_{\alpha \beta}(L=0)$ has only quadratic
dependence, the long-baseline oscillation can have
linear dependence on the RG running parameters
$\beta_X$ \cite{Babu:2021cxe}. For a perturbative
$\beta_X$, the linear dependence would have larger
effect than its quadratic counterpart. However,
longer baseline means significant dilution of the
neutrino fluxes at detector. It is actually of
great advantages to use the short-baseline experiment
for probing the RG running parameters.
For simplicity, we would omit $L = 0$ for the
zero-distance limit and $P_{\alpha \beta}$ as
abbreviation of $P_{\alpha \beta}(L = 0)$
in the later part of this paper.

As we have already noticed, there are four
degrees of freedom in the mixing matrix relevant for oscillation experiments.
We consider the usual parameterization with three mixing angles $\theta_{ij}$ and the Dirac CP phase $\delta_D$, see \cite{Denton:2020igp} for a discussion of the choice of parameterization.
One can map the new physics into RG running of the
four parameters, each with their own $\beta_X$.
Once taking multiple RG running parameters into account, 
degeneracies  can exist among them in the transition
probabilities. For simplicity, we consider only two RG running
parameters at a time in the following discussions which
we find to be sufficient to understand the parameter space.

For a typical short-baseline experiment like the DUNE near detector (DUNE-ND), the relevant channels are $\nu_\mu\rightarrow\nu_e$
and $\nu_\mu\rightarrow\nu_\mu$ as well as their anti-neutrino counterparts. 
In the following, unless otherwise specified, we focus on the discussion about the neutrino mode, while the discussion about the anti-neutrino mode can be derived in a similar way.
When expanded to the quadratic order of $\beta_X$,
the transition probability effective parameters that are involved in the DUNE-ND are,

\begin{subequations}
\begin{align}
   \mathcal B_{\mu e}
&=  
  \beta_{12}^2 c_{13}^2 c_{23}^2  
+ \beta_{12} \beta_{13}  c_\delta c_{13} s_{2\theta_{23}} 
+ \beta_{13}^2 s^2_{23} 
\nonumber\\
&
+ \beta_\delta^2 c_{13}^2 s_{13}^2 s_{23}^2 
- \beta_\delta \beta_{12} c_{13}^2s_\delta s_{13}s_{2\theta_{23}}
+\mathcal O(\beta_{ij}^3)
\label{eq:pme}
\\
 \mathcal B_{\mu\mu}
&= 
-  \beta_{12}^2\left(1 - s_{23}^2c_{13}^2 - s_{13}^2 s^2_\delta s^2_{2\theta_{23}} \right)
- \beta_{13}^2 s^2_{23}
   \nonumber\\
&\quad -\beta_{23}^2 - \beta_\delta^2 s^2_{13}s^2_{23}\left(1-s^2_{13}s^2_{23}\right)
    \nonumber\\
&\quad  - \beta_{12}\beta_{13}c_\delta c_{13}c_{2\theta_{23}} 
-2 \beta_{12}\beta_{23}c_\delta s_{13}
\nonumber\\
&\quad + \beta_\delta\beta_{12}s_\delta s_{13} s_{2\theta_{23}} \left(1-2s^2_{13}s^2_{23}\right)+\mathcal O(\beta_{ij}^3)\,,
\label{eq:pmm}
\end{align}
\label{eq:Pmemm}%
\end{subequations}
where we denote $(c_{ij}, s_{ij}) \equiv (\cos \theta_{ij}, \sin \theta_{ij})$
and $(c_\delta, s_\delta) \equiv (\cos \delta_D, \sin \delta_D)$. 
Here we only keep the RG running parameters up to the second order, while the higher-order terms are neglected. 
As discussed above, the presence of the RG running effect can induce non-trivial transition probabilities.
From the above formula, different transition channels are determined by different RG running parameters.
For example, there is no $\beta_{23}$ term in the expression of $\mathcal{B}_{\mu e}$ while all four $\beta_{ij}$ and $\beta_\delta$ are involved in $\mathcal{B}_{\mu\mu}$.

Nevertheless, the oscillation channels $\nu_\mu\rightarrow\nu_e$
and $\nu_\mu\rightarrow\nu_\mu$ involves all the four RG running
beta functions $\beta_{ij}$ and $\beta_\delta$. Using only
these two channels is not enough to break the degeneracy.
It is then necessary to search for other transition channels to break the degeneracy.

However, not all oscillation probabilities are independent.
Although the unitarity property for a single transition
probability has been lost \cite{Ge:2024ibn}, the sum
of transition probabilities still preserve some unitarity
properties,
\begin{align}
  \sum_\alpha P_{\alpha\beta}(E_\nu; Q^2_p, Q^2_d)
=
  \sum_\beta P_{\alpha \beta}(E_\nu; Q^2_p, Q^2_d)
=
 1.
\label{eq:unitarity-sum}
\end{align}
In the zero-distance limit, the
transition probabilities only involve the mixing
matrices as shown in \geqn{eq:prob} and the summation
above can then be expressed as,
\begin{align}
  \sum_\alpha P_{\alpha \beta}
=
  \left[ U_p U^\dagger_d U_d U^\dagger_p \right]_{\beta \beta}
=
  \delta_{\beta \beta}.
\end{align}
Even with nonzero baseline, the evolution
phase sandwiched between the two mixing matrices
would not change the final conclusion.
This unitarity relation shows that among the
total nine transition channels, only four
transition channels are independent.

Taking the unitarity condition in \geqn{eq:unitarity-sum}
for the probability sum into account, only 4 of the
9 oscillation probabilities could be independent.
Besides the aforementioned $\nu_\mu \rightarrow \nu_e$
and $\nu_\mu \rightarrow \nu_\mu$ at the near detector
of the DUNE accelerator experiment, one may also
consider the JUNO-TAO near detector at the reactor
experiment whose transition probability is determined by
\begin{align}
  \mathcal B_{ee}
=
- \beta_{13}^2
- \beta_{12}^2c^2_{13}
- \beta_\delta^2 c_{13}^2s_{13}^2.
\label{eq:Pee}
\end{align}

In addition, we take the FASER$\nu$2 experiment that
consists of $\mathcal{O}$(TeV) neutrinos 
into account. Due to its high-energy and 
multiple neutrino sources with three flavors produced, FASER$\nu$2 can 
detect all three flavor neutrinos.  
Particularly interesting is the $P_{e\tau}$
transition probability with,
\begin{align}
   \mathcal B_{e \tau}
& =
   \beta_{12}^2 c_{13}^2 s_{23}^2 
-  \beta_{12} \beta_{13} c_{\delta} c_{13} s_{2\theta_{23}} 
+  \beta_{13}^2 c_{23}^2.
\end{align}

Putting things together, we may see that $\mathcal B_{\mu e}$
in \geqn{eq:pme}, $\mathcal B_{ee}$ in \geqn{eq:Pee},
and $\mathcal B_{e \tau}$ above share the same set of beta
functions ($\beta_{12}$, $\beta_{13}$, and $\beta_\delta$).
Such combination of three independent oscillation channels
can really constrain the three beta functions. Then the
remaining $\mathcal B_{\mu \mu}$ can constrain the fourth
beta function $\beta_{23}$.

As multiple RG running parameters are simultaneously involved,
certain degeneracy among these RG running parameters exist
in any single channel. Explicitly, $\mathcal B_{\mu e}$
has a cross term of $\beta_{12} \beta_{13}$ which
indicates a correlation between them.
Notice that the correlation depends on the sign of $c_\delta$. 
Namely, since both prefactors of $\beta_{12}^2$ and $\beta_{13}^2$ are positive in $\mathcal{B}_{\mu e}$, 
there exists a positive (negative) correlation between $\beta_{12}$ and $\beta_{13}$ for $c_\delta<0$ ($c_\delta>0$).
In the case that $c_\delta=0$, there is no correlation between  $\beta_{12}$ and $\beta_{13}$.
Similarly, although the cross term of $\beta_{12}$ 
and $\beta_{13}$ in $\mathcal{B}_{\mu\mu}$ has an 
opposite sign to its counterpart in $\mathcal{B}_{\mu e}$, 
the prefactors of $\beta_{12}^2$ and $\beta_{13}^2$ are also opposite. Consequently,
the correlation behavior between $\beta_{12}$ and $\beta_{13}$ in $\mathcal{B}_{\mu\mu}$ is the same as the one in $\mathcal{B}_{\mu e}$.
Below we will show with more details how the three experiments
(DUNE-ND, JUNO-TAO, and FASER$\nu$2) can break the degeneracy
among the RG running beta functions.

\begin{figure}[t!]
\centering
\includegraphics[width=0.48\textwidth]{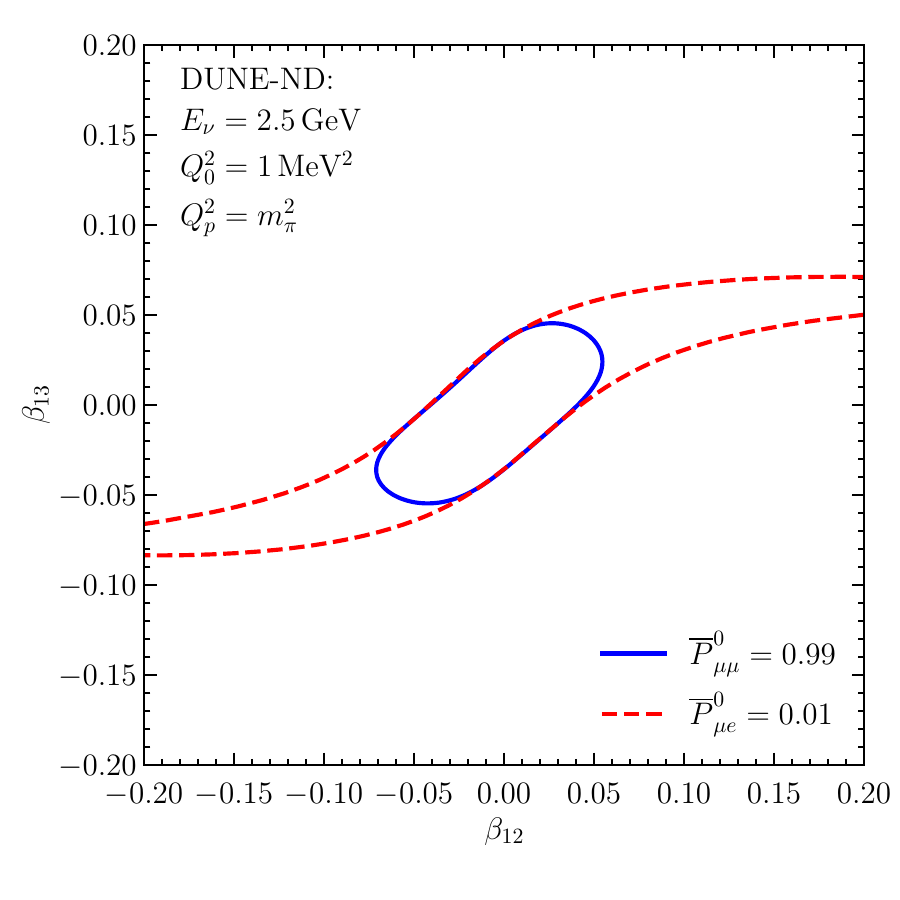}
\caption{The momentum transfer averaged transition probabilities $\overline{P}^0_{\mu\mu} = 0.99$ (blue solid) and
$\overline{P}^0_{\mu e} = 0.01$ (red dashed)
on the $\beta_{12}$--$\beta_{13}$ space. The neutrino energy $E_\nu$ is set
as 2.5\,GeV and the new physical scale $Q^2_0$ is set as 1\, MeV$^2$.
The production momentum transfer is fixed at $m_\pi^2$. We take the global-fit result 
of $\delta_D = 194^\circ$ from \cite{deSalas:2020pgw}.}
\label{fig:prob-dune}
\end{figure}

\section{Breaking Degeneracies with Low- and High-energy SBL Experiments}
\label{sec:degeneracies}
As shown above, degeneracies exist among multiple RG running
parameters especially when considering only a single 
experiment at a time.
This will lead to a correlation between $\beta_{12}$ 
and $\beta_{13}$ in both detection 
channels, namely, the $\nu_\mu$ 
disappearance channel and $\nu_e$ appearance 
channel. Such a degeneracy can significantly reduce the experimental sensitivity to the multiple RG running parameters when only one experiment is considered at a time.
To numerically calculate the transition probability and show the correlation behavior for a realistic experiment, we
take a momentum transfer averaged transition probability,
\begin{equation}
    \overline{P^0_{\alpha\beta}}
\equiv 
    \int_{Q^2_{d,\rm min}}^{Q^2_{d,\rm max}}
    P^0_{\alpha\beta}(Q^2_d)
    \left[\frac{1}{\sigma}\frac{d\sigma}{d Q^2_d}\right]
     d Q^2_d\, .
\end{equation}
where $d\sigma/dQ^2_d$ is the differential detection cross 
section and $\sigma$ is the total cross section.
We use this approach since
in most neutrino experiments the detection momentum transfer 
is not fixed, but follows a distribution determined
by the neutrino energy and detection cross-section.
However, neutrinos are produced from particle decays via the charged 
current weak interaction and hence the production momentum transfer is 
constant and equal to the parent particle mass squared.

\gfig{fig:prob-dune} 
shows the averaged probabilities $\overline{P^0_{\mu\mu}}$ 
(blue) and $\overline{P^0_{\mu e}}$ (red) that give a 1\% correction to the standard picture, on the 
$\beta_{12}$--$\beta_{13}$ space. For illustration,
we take the neutrino peak energy $E_\nu = 2.5$\,GeV
of the DUNE experiment
and $Q^2_p = m_\pi^2$ \cite{DUNE:2020ypp}. 
We also take the global-fit result $\delta_D = 194^\circ$ 
\cite{deSalas:2020pgw} and the detection momentum distribution
was obtained with GENIE \cite{Andreopoulos:2009rq}.
In both curves, there is a positive correlation between
$\beta_{12}$ and $\beta_{13}$, which is 
consistent with the analytical formula of \geqn{eq:Pmemm}.
The blue curve has an ellipse-like shape with the major
axis directed towards the 1st quadrant 
direction. The red curve matches the blue 
curve for $|\beta_{ij}| \lesssim 0.05$. While for $|\beta_{ij}| > 0.05$ 
the red curve has an elongated tail extending above
$|\beta_{ij}| > 0.2$ which comes from the $\beta_X^4$ terms 
omitted in \geqn{eq:pme} and \geqn{eq:pmm}.
The existence of the positive correlation
generates the degeneracy between the RG running parameters
which can significantly reduce the experimental sensitivity. 

\begin{figure}[t!]
\centering
\includegraphics[width=0.48\textwidth]{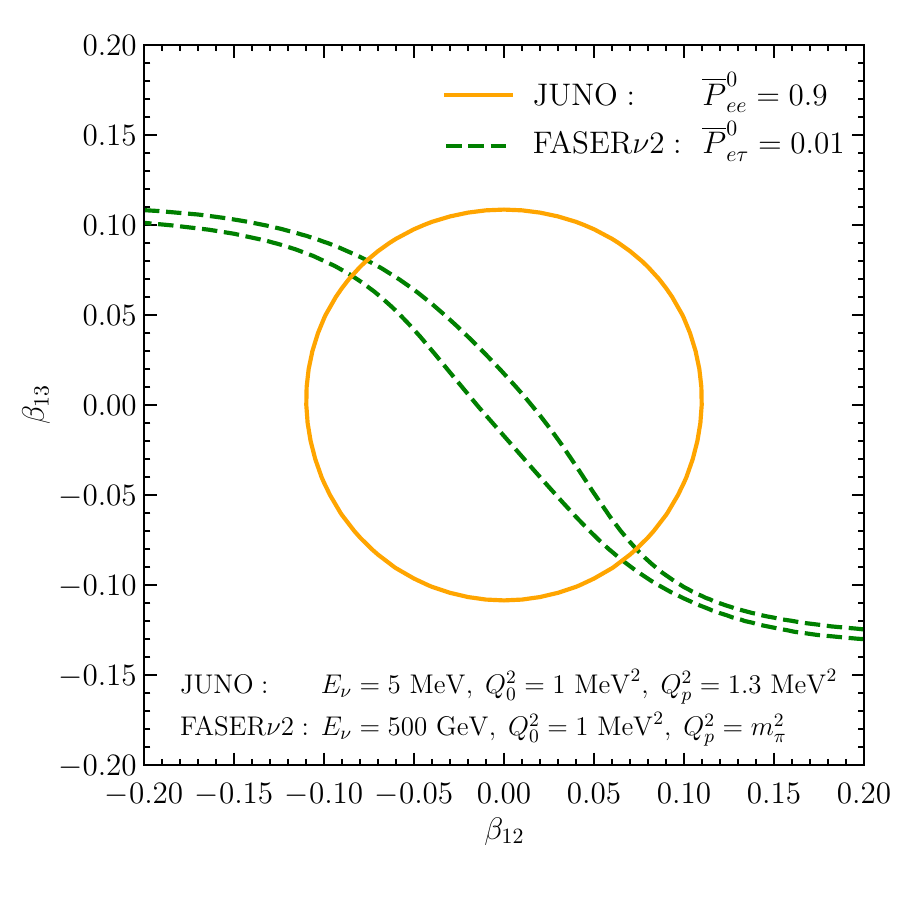}
\caption{The averaged transition probability
$\overline{P}^0_{ee} = 0.9$ at JUNO (orange solid)
and $\overline{P}^0_{e \tau} = 0.01$ at FASER$\nu$2 (green dashed)
on the $\beta_{12}$--$\beta_{13}$ space. The neutrino energy $E_\nu$ is set
as 5\,MeV and the new physical scale $Q^2_0$ is set as 1\,MeV$^2$.
The production momentum transfer is fixed as $(m_n-m_p)^2\approx 1.3\,$MeV$^2$.}
\label{fig:prob-faser-juno}
\end{figure}

In general, there are 
two possible ways to look for complementary experiments involved with different transition 
channels from DUNE-ND. One is using different neutrino sources. 
As mentioned above, the major neutrino flux in the production at DUNE-ND is muon flavor. 
Therefore, the possible different neutrino 
sources are electron and tau flavors. 
However, due to experimental challenges, it is difficult to produce a 
sufficient tau-flavor neutrino flux to have a large 
statistics. Therefore, it is promising to 
focus on the electron-flavor flux.

Indeed, the
$\bar\nu_e$ flux can be sufficiently 
produced from a reactor and can be detected 
by the near detector in a typical reactor experiment. In this paper, we consider
JUNO-TAO \cite{JUNO:2020ijm} as a benchmark experiment. Especially, the corresponding
transition probability for the reactor neutrino experiment
has been summarized in \geqn{eq:Pee}.
Being different from \geqn{eq:Pmemm}, there is no 
cross term among the RG running parameters. Therefore, no 
correlation between $\beta_{12}$ and $\beta_{13}$ 
appears in the electron disappearance channel.

The orange contour in \gfig{fig:prob-faser-juno}
on the $\beta_{12}$--$\beta_{13}$ space shows the 
transition probability $\bar{P}_{ee} = 0.9$ with 
typical neutrino energy $E_\nu=5\,$MeV and 
production momentum transfer 
$Q^2_p=1.3\,$MeV$^2$ at JUNO-TAO \cite{Ge:2024ibn}. 
The contour is a circle, indicating that 
there is no correlation between 
$\beta_{12}$ and $\beta_{13}$ which is consistent with
\geqn{eq:Pee}. This feature 
provides JUNO-TAO as a powerful platform
to constrain on $\beta_{12}$ and 
$\beta_{13}$ without any contamination from the degeneracy.

Besides the different neutrino 
flavor source from production, the other 
practical way for breaking degeneracies is 
expanding the detection possibility. Typically,
the detection channels are limited by the energy threshold. 
Explicitly, the JUNO-TAO can only detect electron flavor since its $\mathcal{O}$(MeV) neutrino energies are below the muon or tau mass. 
Similarly, with $\mathcal{O}$(GeV) 
energies, DUNE-ND can only efficiently 
detect electron and muon flavors, although 
a small number of tau events can be 
measured with high uncertainties.  
Therefore, in order to expand the detection possibility, we need to consider other higher-energy experiments.

As illustrated in \gfig{fig:prob-faser-juno},
we have also shown the transition
probability $\overline{P^0_{e\tau}}= 0.01$ for the electron flavor
transition into the tau flavor. Here we take $500\,$GeV neutrino energy
and $m_\pi^2$ production momentum transfer as representative values. Unlike the positive
correlation observed in \gfig{fig:prob-dune},
$\overline{P^0_{e\tau}}$ has a negative correlation.
The combination of measurements with oppositely signed correlations allows for the removal of the degeneracy.
More intuitively, although the overlapping area of the contours in 
\gfig{fig:prob-faser-juno} and \gfig{fig:prob-dune} remains,
it is much smaller when compared with each region separately.

To further illustrate the transition probabilities in the three experiments mentioned above, 
we define the average of the logarithm term $\overline{\ln^2|Q^2_d|/|Q^2_p|}$ 
to obtain the 
detection-averaged probability,
\begin{align}
  \overline{P^0_{\alpha\beta}}
\equiv
   \delta_{\alpha \beta} 
+  \mathcal B_{\alpha \beta} ~~
   \overline{\ln^2\left(\frac{|Q^2_d|}{|Q^2_p|}\right)}.
\end{align}
The transition channels and values of the average logarithm squared for JUNO-TAO 
\cite{JUNO:2020ijm}, DUNE-ND\cite{DUNE:2021tad}, and 
FASER$\nu$2 \cite{FASER:2024ykc} are summarized in 
\gtab{tab:exp}. The $\ln^2$ terms are 
$\mathcal O(10)$, meaning that a $\beta_X \sim \mathcal 
O(0.1)$ is close to the non-linear regime. In the non-linear
regime, large changes to oscillation parameters take place and 
the oscillation data needs to be carefully re-analyzed. Moreover, 
due to the correlations described above, constraining the 
$\beta_X$ parameters within the linear regime is a difficult 
task for any experiment alone. Fortunately, as we will 
show, the combination of the three future experiments 
JUNO-TAO, DUNE-ND and FASER$\nu$2 is ideal to bring down 
the bounds on the RG running of neutrino mixing parameters
to the linear regime.

\begin{table}[t!]
\centering
\begin{tabular}{c|| c | c}
Exp. &   Oscillation Channels & $\overline{\ln^2|Q^2_d|/|Q^2_p|}$
\\
\hline
\hline
JUNO-TAO
& 
$\overline\nu_e\rightarrow\overline\nu_e$
& \makecell[c]{ 5 \\ ($E_\nu = 3.6$\,MeV)}
\\
\hline
DUNE-ND
& 
\makecell[c]{
$\nu_{\mu}\rightarrow \nu_{e,\mu}$ ~\&~ $\overline\nu_{\mu}\rightarrow \overline \nu_{e,\mu}$\\
$\nu_{e}\rightarrow \nu_{e,\mu}$ ~\&~ $\overline\nu_{e}\rightarrow \overline \nu_{e,\mu}$
} &  \makecell[c]{ 12 \\ ($E_\nu = 2.5$\,GeV)}
\\
\hline
FASER$\nu$2 
&
\makecell[c]{
$\nu_{\mu}\rightarrow \nu_{e,\mu,\tau}$ ~\&~ $\overline\nu_{\mu}\rightarrow \overline \nu_{e,\mu,\tau}$\\
$\nu_{e}\rightarrow \nu_{e,\mu,\tau}$ ~\&~ $\overline\nu_{e}\rightarrow \overline \nu_{e,\mu,\tau}$\\
$\nu_{\tau}\rightarrow \nu_{e,\mu,\tau}$ ~\&~ $\overline\nu_{\tau}\rightarrow \overline \nu_{e,\mu,\tau}$\\
}
&  \makecell[c]{ 82\\ ($E_\nu = 0.5$\,TeV)}
\\
\hline
\end{tabular}
\caption{Oscillation channels at JUNO-TAO, DUNE-ND, and FASER$\nu$2 and the detection-averaged squared logarithm for
a representative neutrino energy $E_\nu$.}
\label{tab:exp}
\end{table}

\section{Sensitivities to RG Running Parameters}
\label{sec:sensitivities}

To optimize the sensitivity of the RG parameters,
we combine three SBL experiments: JUNO-TAO, DUNE-ND, and 
FASER$\nu$2, which cover a large energy range from MeV up to 
TeV and provide independent measurement of different 
transition channels: $P^0_{ e 
 e}$, $P^0_{\mu e}$, and $P^0_{ e \tau}$, respectively. 
As shown above, these transition channels have totally different correlation behavior.
For completeness, the 
transition channels of these three experiments are summarized 
in \gtab{tab:exp}.

As the neutrino production and detection processes are different among these three experiments, we need different considerations to calculate the predicted event rate.
Since JUNO-TAO is already taking data \cite{JUNO:2025gmd} and its systematics is established \cite{JUNO:2020ijm}, we use the complete event distribution over the reconstructed energy and include the systematic in our simulations.
For the future DUNE-ND and FASER$\nu$2 detectors, we take a more conservative approach and consider only the total number of events expected to be observed in each experiment. Even with such crude observable, the sensitivity obtained reaches competitive levels, especially when combining different experiments.

For the JUNO-TAO experiment, the initial neutrino flux contains only electron 
antineutrinos. Also, because the neutrino energies 
are below the muon or tau mass, only $\bar\nu_e$ can 
be measured at the detector. The predicted true 
neutrino event rate contains only a single channel that can be calculated,
\begin{equation}
    \frac{\dd N^{\rm JUNO-TAO}_{\bar\nu_e\rightarrow\bar\nu_e}}{\dd E_\nu}
    =
    N_{t} 
    \Phi_{\bar\nu_e} 
    \overline{P^0_{e e}}
    \sigma_{\bar\nu_e},
\label{eq:Events_JUNO}
\end{equation}
where $N_t$ is the normalization factor, $\Phi_{\bar\nu_e}$ the initial 
$\bar\nu_e$ flux, which is a combination of several reactors
each containing different fractions of isotope.
Also, the detection cross section $\sigma_{\bar\nu_e}$
is taken to be the inverse beta decay cross section.
In order to simulate the energy resolution of the
detector, we implement a Gaussian smearing function
with energy resolution of $1.5\%/\sqrt{E / {\rm MeV}}$ \cite{JUNO:2020ijm} 
where $E \equiv E_{\bar{\nu}_e} - 0.8\,$MeV
is the deposit energy.

The quantity  $\overline{P^0_{e e}}$ is 
the zero-distance momentum transfer averaged
transition probability for $\bar\nu_e\rightarrow\bar\nu_e$.
In order to calculate the transition probability, 
one needs three momentum transfer scales, $Q^2_0$,
$Q^2_p$, and $Q^2_d$. For illustration, we take $Q^2_0=1\,$MeV$^2$ throughout this paper. We also take $Q^2_p=(m_n-m_p)^2$ corresponding to the beta decay momentum transfer and an analytical distribution function for $Q^2_d$ \cite{Giunti:2007ry}. The backgrounds come from  accidental, fast neutron, and 
$^{9}$Li/$^{8}$He backgrounds and
their spectra are extracted from \cite{JUNO:2024jaw}.
The sensitivity is calculated by a sum over all the bins through the function $\chi^2_{\rm JUNO-TAO}$,
\begin{align}
\hspace{-.3cm}
  \chi^2_{\rm JUNO-TAO}
& \equiv 
  \sum_i^{\rm bins}
  \left(\frac{N_{i}^{\rm true} - N_{i}^{\rm test}
      }{\sqrt{N_{i}^{\rm true}}} \right)^2
+ \chi_{\rm JUNO-SYS}^2.
\label{eq:chi_JUNO}
\end{align}
The systematic uncertainties in $\chi_{\rm JUNO-SYS}^2$ includes several normalization and tilt 
parameters \cite{JUNO:2020ijm}. For a 
more detailed description of our JUNO-TAO simulation 
refer to \cite{Ge:2024ibn}. 
To derive the sensitivity, we assume 6.5 years of data-taking time.

For DUNE-ND, we take only the ND-LAr module \cite{DUNE:2021tad} into account. 
There are two running modes at DUNE, namely, the neutrino mode and
anti-neutrino mode. 
In each mode, there are four neutrino components in the initial 
neutrino flux: $\nu_\mu, \bar\nu_\mu,\nu_e$, and $\bar\nu_e$ \cite{DUNE:2020ypp}.
As for neutrino detection, we follow the DUNE near detector CDR \cite{DUNE:2021tad}
and classify them into two categories.
For the neutrino mode, they are $\nu_\mu$ and $\nu_e+\bar\nu_e$, 
while for the anti-neutrino mode, they are $\bar\nu_\mu$ and $\nu_e+\bar\nu_e$.
Therefore, the predicted event rates in the neutrino mode are,
\begin{subequations}
\begin{align}
    \frac{\dd N^{\rm DUNE-ND}_{\nu_\alpha\rightarrow \nu_\mu}}{\dd E_\nu}
& \equiv 
    N_{t} \Phi_{\nu_\alpha}
    \overline{P^0_{\alpha \mu}}\sigma_{\nu_\mu}\,,
\\
    \frac{\dd N^{\rm DUNE-ND}_{\nu_\alpha\rightarrow \nu_e}}{\dd E_\nu}
& \equiv 
    N_{t}
    \left[
        \Phi_{\nu_\alpha}
    \overline{P^0_{\alpha e}}\sigma_{\nu_e}
    +   \Phi_{\bar\nu_\alpha}
    \overline{P^0_{\bar\alpha \bar e}}
    \sigma_{\bar \nu_e}
    \right]  \,,
\end{align}
\label{eq:Events_DUNE}%
\end{subequations}
and the predicted event rates in the 
anti-neutrino mode can be calculated in a similar way.

Since most DUNE neutrinos are produced from the decay of pions and muons which have a similar mass,
we fix the neutrino production momentum transfer 
at $Q^2_p=m_\pi^2$ for simplicity.
For the detection momentum transfer, we take the distribution 
from the GENIE \cite{Andreopoulos:2009rq} simulation results. 
To be conservative, we don't include the energy spectrum information for the DUNE-ND, but only the total number of events for each mode and channel. Moreover, we include an overall nuisance parameter for each flux source when calculating the experimental sensitivity. 
For example, the muon-neutrino detection channel in the neutrino mode has,
\begin{align}
\hspace{-3mm}
  \chi^2_{\nu\text{-mode}, \nu_\mu}
\equiv 
  \left(\frac{\sum_\alpha (1+a_\alpha^{\nu\text{-mode}})N^{\rm RG}_{\nu_\alpha\rightarrow\nu_\mu}
- N^{\rm std}_{\nu_{\mu}}}{\sqrt{N^{\rm std}_{\nu_{\mu}}}}\right)^2,
\end{align}
where $N^{\rm RG}_{\nu_\alpha\rightarrow\nu_\mu}$
is the bin-summed prediction according to \geqn{eq:Events_DUNE}
with the RG running effect and $N^{\rm std}_{\nu_{\mu}}$
is the total predicted 
number of muon neutrino events in the standard oscillation scenario. Here $a_\alpha^{\nu\text{-mode}}$ is the overall nuisance parameter of $\alpha$ in the neutrino mode. 
Summing up two neutrino modes and all relevant channels,
the total DUNE-ND $\chi^2$ function is,
\begin{align}
    \chi^2_{\rm  DUNE\!-\!ND}
=
 \sum_{\substack{\rm mode,}\beta}
   \chi^2_{\rm mode, \beta}
  +\sum_{\substack{\rm mode,}\alpha}
   \left(\frac{a^{\rm mode}_{\alpha}}{\sigma_{a^{\rm mode}_\alpha}}\right)^2\,.
\label{eq:chi_DUNE}
\end{align}
Here $\beta\in \{\nu_\mu, \nu_e+\bar\nu_e\}/\{\bar\nu_\mu, \nu_e+\bar\nu_e\}$ denotes the observable in the neutrino/anti-neutrino mode and $\alpha\in\{\nu_\mu,\bar\nu_\mu,\nu_e,\bar\nu_e\}$ denotes the neutrino source.
Each nuisance parameter $a_\alpha^{\rm mode}$ is minimized over with a Gaussian prior with uncertainty $\sigma_{\alpha}^{\rm mode} = 10\%$.

For FASER$\nu$2, the neutrinos in both production and detection processes can be classified into three categories: $\nu_e+\bar\nu_e$, $\nu_\mu+\bar\nu_\mu$, and $\nu_\tau+\bar\nu_\tau$ \cite{FASER:2024ykc}.
Compared with the above two experiments, the neutrino production process
is more complicated at the FASER$\nu$2 experiment. 
Explicitly, neutrinos are produced from the decay of multiple 
particles which have different masses with comparable contributions. Recall that the 
neutrino production momentum transfer is equal to the mass squared of the parent particle. Consequently,
the calculation of event rates at FASER$\nu$2 should be divided into contributions from various parent particles and then summed up,
\begin{align}
    \frac{\dd N^{{\rm FASER}\nu2}_{\nu_\alpha \rightarrow\nu_\beta}}{\dd E_\nu}
&=
    N_{t}
    \sum_{\rm parents}
    \left[
    f_{\rm p}
    \Phi_{\nu_\alpha}
    \overline{P^{0,~p}_{\alpha \beta}}
    \sigma_{\nu_\beta}
    \right. 
\nonumber 
\\ & \qquad \qquad \qquad 
\left. 
+  f_{\bar{\rm p}}
    \Phi_{\bar\nu_\alpha}
    \overline{P^{0,~p}_{\bar\alpha \bar\beta}}
    \sigma_{\bar\nu_\beta}
    \right],
\label{eq:faserN}
\end{align}
where $f_{\rm p}$ is the fraction of a given parent particle among all parents and $\overline{P^{0,~p}_{\bar\alpha \bar\beta}}$ is the 
momentum transfer averaged probability calculated with the production momentum transfer at the parent particle mass, $Q^2_p = m_p^2$.
The main neutrino parent particles in the electron/muon/tau
flavor production are (kaon, Dmeson, hyperon)/(pion, kaon,
Dmeson)/(Dmeson) \cite{FASER:2024ykc}, respectively. Moreover,
the fraction of each source depends on the neutrino energies \cite{FASER:2024ykc}.
For the detection momentum 
transfer, we still take its distribution from the GENIE
\cite{Andreopoulos:2009rq} simulation.

The $\chi^2$ for the detection of $\nu_\beta$ at FASER$\nu$2 is constructed similarly to the DUNE-ND, by summing over all the energy bins and introducing an overall nuisance parameter for each neutrino source $\alpha$,
\begin{align}
    \chi^2_{\beta}
& \equiv 
    \left(\frac{\sum_\alpha (1+a_\alpha)N^{\rm RG}_{\alpha \beta}
    - N^{\rm std}_{\beta}}{\sqrt{ N^{\rm std}_{\beta}}}\right)^2.
\end{align}
Similarly, $N^{\rm RG}_{\alpha \beta}$ ($N^{\rm std}_{\nu_{\mu}}$ ) is the bin-summed prediction with RG running effect (standard oscillation) using \geqn{eq:faserN}.
Summing up the relevant channels, the total $\chi^2$ for FASER$\nu$2 is,
\begin{align}
    \chi^2_{\rm  FASER\nu2}
=
 \sum_{\substack{\beta}}
    \chi^2_{\beta}
    +
    \sum_\alpha \left(\frac{a_{\alpha}}{\sigma_{a_\alpha}}\right)^2.
\label{eq:chi_FASER}
\end{align}
There are three nuisance parameters $a_\alpha$,  $\alpha = e, \mu$ and $\tau$ corresponding to a Gaussian prior with uncertainty $\sigma_{e} = \sigma_{\mu} =50\%$, and $ \sigma_{\tau} = 100\%$, which is more conservative than the values used in the literature \cite{Ansarifard:2021dju,Celestino-Ramirez:2023zox}.

The predicted event rates in \geqn{eq:Events_JUNO},
\geqn{eq:Events_DUNE}, and \geqn{eq:faserN} give the event
distribution as a function of neutrino energy. In the 
presence of the RG running effect, a non-zero 
transition probability will lead to 
a different event rate from the standard case
even at the total number of event level. Therefore, 
the short-baseline neutrino experiments provide
a unique platform to test the 
RG running effect by comparing the measured event rates with 
the predicted one. The sensitivity should improve with a full 
energy spectrum analysis and also if the detection momentum 
transfer can be measured \cite{Ge:2023azz}. However, such 
analysis requires complicated treatment
of the systematics, for the bin-to-bin uncertainties,
which is beyond the scope of this work.
Therefore, to simplify the calculation of the experimental 
sensitivity, we focus on the total event numbers for DUNE-ND and FASER$\nu$2 and include the energy spectrum information only for JUNO-TAO.

\begin{figure}[t!]
\centering
\includegraphics[width=0.48\textwidth]{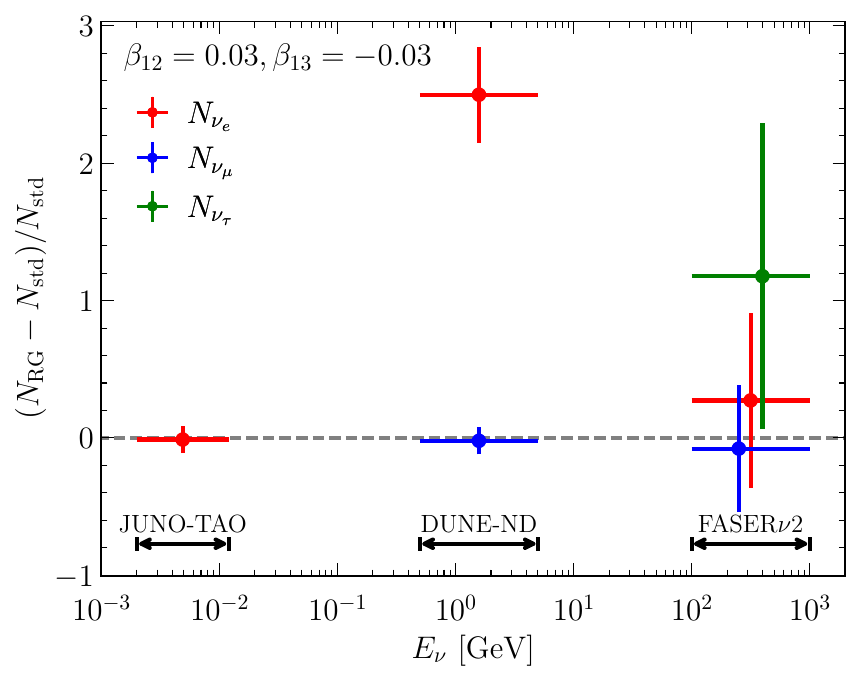}
\caption{The total event difference fraction of flavor $\nu_\alpha$ between the standard ($N_{\alpha, \rm std}$) oscillation and with RG-running effect ($N_{\alpha, \rm RG}$) for JUNO-TAO, DUNE-ND and FASER$\nu$2 experiments. For illustration, we take the only non-zero RG-running parameters to be $\beta_{12} = 0.03$ and $\beta_{13} = - 0.03$. The error in the $x$-direction is estimated by taking the full energy range of each experiment, while the $y$-direction error is estimated using Eq.\,\eqref{eq:ErrorEstimate}.}
\label{fig:NRG_Nstd}
\end{figure}

\begin{figure}[t] 
\centering
\includegraphics[width=0.48\textwidth]{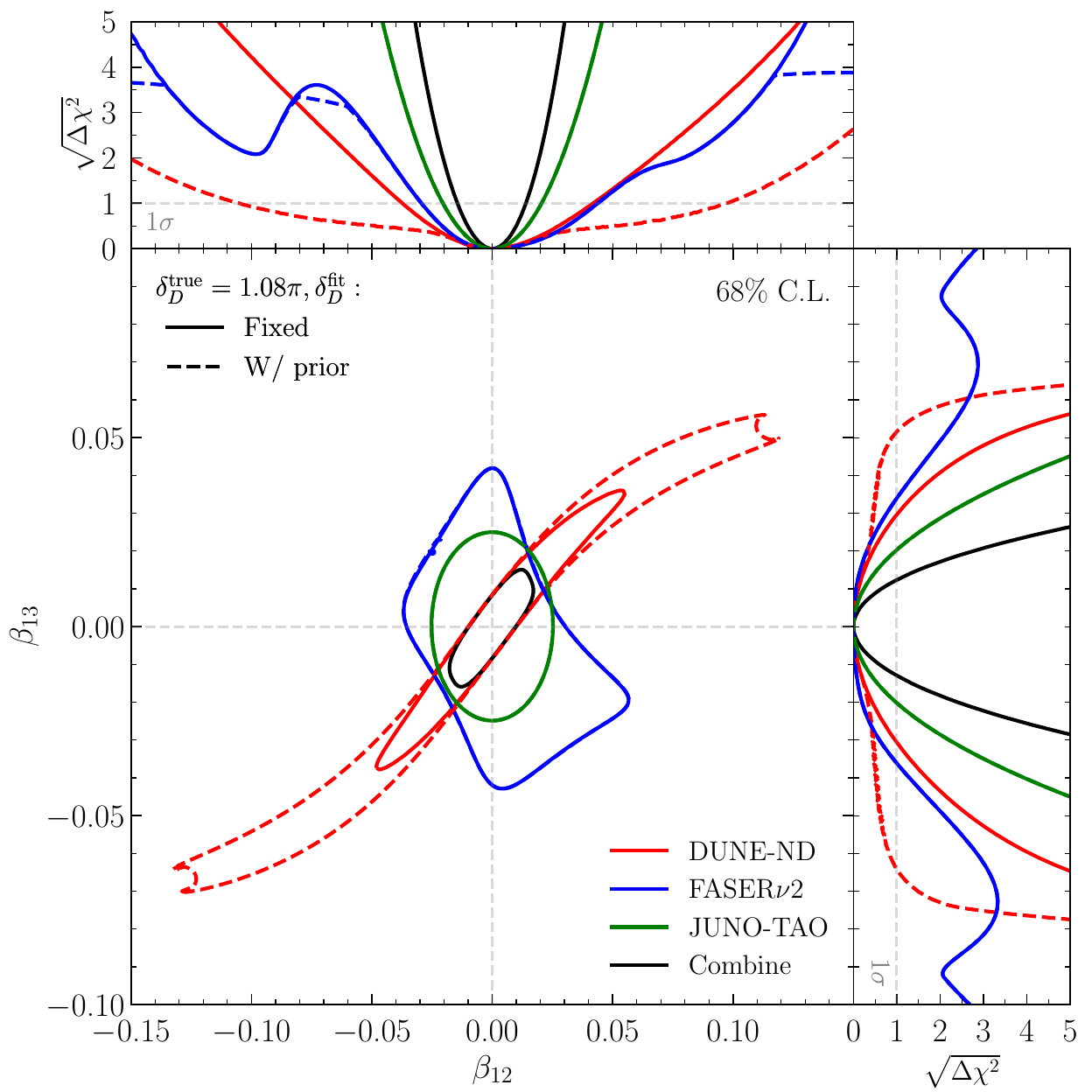}
\caption{The JUNO-TAO (green), DUNE-ND (red), and FASER$\nu$2 
(blue) expected sensitivity to the $\beta_{12}$ and 
$\beta_{13}$ RG running parameters at 68\% C. L. The 
solid lines represents fixed at the best-fit value of  
\cite{deSalas:2020pgw} ($\delta_D = 1.08\pi$) while the 
dashed lines minimize over the value of $\delta_D$ with prior knowledge on its uncertainty. The black 
curve represents the combination of all the three experiments sensitivities and represents a significant improvement on the sensitivity when compared with the separated cases
}
\label{fig:beta12_vrs_beta13}
\end{figure}

For illustrating the power of total event number analysis, in 
\gfig{fig:NRG_Nstd} we show the expected event difference 
fraction, $\Delta N_{\beta}/N_{\beta}^{\rm std} \equiv 
(N_{\beta}^{\rm RG}-N_{\beta}^{\rm std})/N_{\beta}^{\rm std}$ of $\beta$
flavor detection for the experiments JUNO, DUNE and 
FASER$\nu$2. When displaying the uncertainty in the y-axis 
direction, we combine the statistical and systematic ones. For 
example, FASER$\nu$2 has $\mathcal O(10^6)$ $\nu_\mu$ events, 
which makes the statistical error tiny $\Delta N^{\rm 
error}_\mu \approx \sqrt{N_{\mu}^{\rm std}}/N_{\mu}^{\rm std} 
\sim 0.05\%$. On the other side, the systematical error for 
each flux is of the order of $50\%$ to $100\%$, however, the 
overall error is reduced due to the correlation between the 
different measuring channels. In order to take into account 
some of these features, we make an estimate about the error 
using the diagonal of a covariance matrix $\Delta 
N_{\beta}^{\rm error} \approx \sqrt{(\Sigma^{-1})_{\beta 
\beta}}$ such that
\begin{equation}
 \Sigma_{\beta\alpha }
\equiv 
    \frac{\delta_{\alpha \beta}}{N_{\beta}^{\rm std}}
- \frac{
    ({\bf N}_{\beta}^{\rm RG}/ N_{\beta}^{\rm std})
    ( {\bf N}_{\alpha}^{\rm RG}/ N_{\alpha}^{\rm std})
   }{ 
   \sum_\gamma ({\bf N}_{\gamma}^{\rm RG})^2 (N_{\gamma}^{\rm std})^{-1}
 +  \sigma^{-2} }\,,
\label{eq:ErrorEstimate}
\end{equation}
where $\sigma$ is the normalization error. Notice that the actual 
error and correlations included in our sensitive analysis is 
more complicated than the one in the equation above and provides 
better sensitivity. In any case, the error bars displayed in 
\gfig{fig:NRG_Nstd} provide a good estimate on the 
expected results. In fact we clearly see that the $\nu_e$ 
channel (red) for DUNE-ND and the $\tau$ channel (green) for 
FASER$\nu$2 can provide a powerful sensitivity to the RG effects 
at the total event rate level since the gray-dashed line 
representing zero difference from the standard case is below 
the error bars. 

\begin{figure}[t]
\centering
\includegraphics[width=0.48\textwidth]{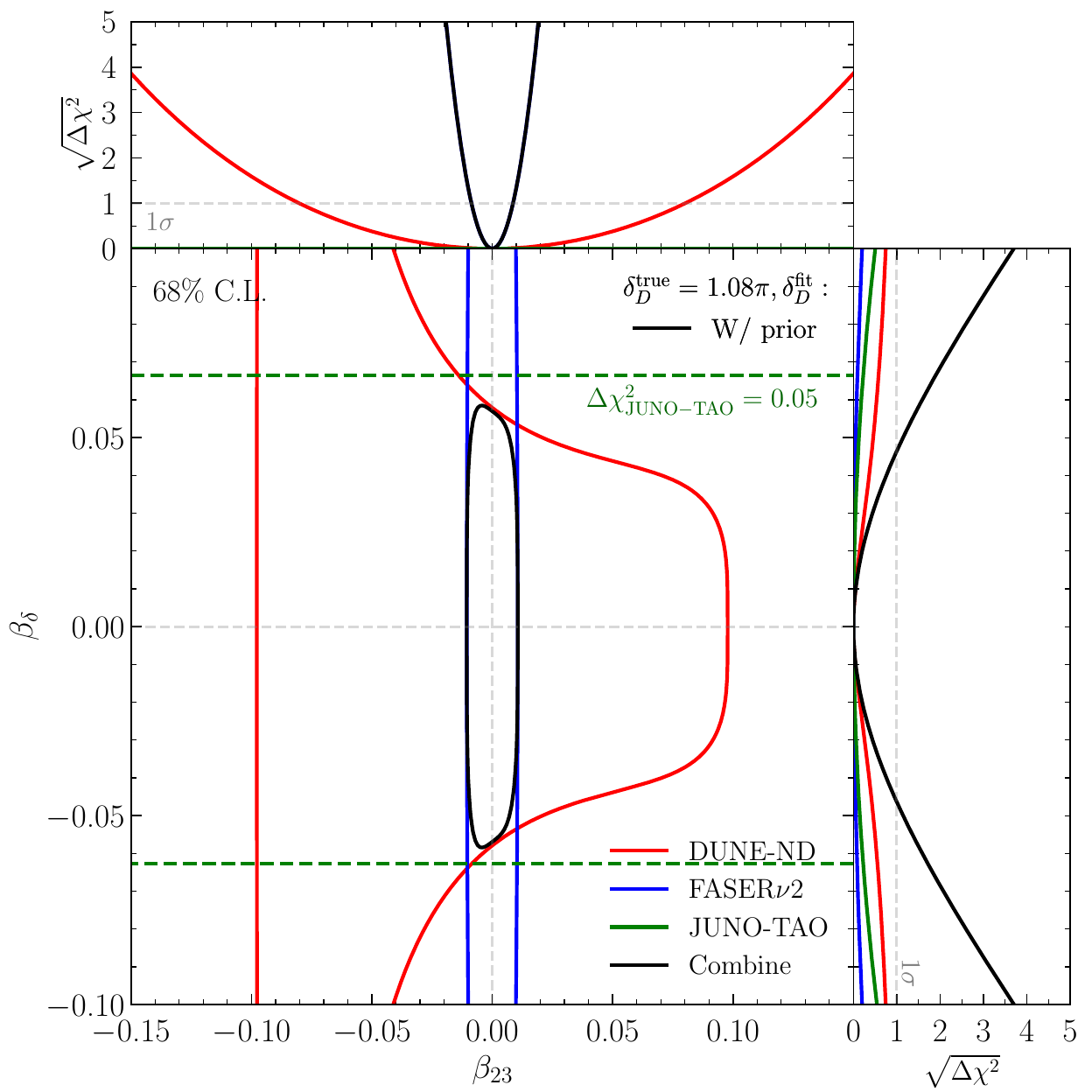}
\caption{The JUNO-TAO (dashed-green), DUNE-ND (red), and 
FASER$\nu$2 (blue) expected sensitivity to the $\beta_{23}$ 
and $\beta_{\delta}$ renormalization parameters at 68\% C. L. 
The JUNO-TAO experiment is insensitive to the $\beta_{23}$ parameter, so the region is obtained by the 1D interval obtained by the condition $\Delta \chi^2 = 0.05$, while the DUNE-ND and FASER$\nu$2 are sensitive to both parameters and the curves are obtained with the 2D confidence region obtained from $\Delta \chi^2 = 2.3$. In all cases we fix $\delta_D = 1.08\pi$. The combined sensitivity (black) represents a significant improvement on the sensitivity when compared with the separated cases.}
\label{fig:beta23_vrs_betadelta}
\end{figure}

Our sensitivity analysis provides the separate sensitivity for JUNO-TAO, DUNE-ND, and FASER$\nu$2 according to the $\chi^2$ 
described by \geqn{eq:chi_JUNO}, \geqn{eq:chi_DUNE},
and \geqn{eq:chi_FASER}, respectively.
For the combined analysis, we adopt the total $\chi^2$ accordingly,
\begin{align}
    \chi^2
\equiv 
  \chi^2_{\rm TAO}
+ \chi^2_{\rm DUNE-ND}
+ \chi^2_{{\rm FASER}\nu2}
+ \chi^2_{\rm prior}.
\end{align}
The last term $\chi^2_{\rm prior}$ includes a Gaussian prior for the Dirac CP phase $\delta_{\rm D}$ phase whose uncertainty is taken from the global-fit result \cite{deSalas:2020pgw}. All the other neutrino oscillation parameters are taken at their best-fit values \cite{deSalas:2020pgw}.

The sensitivity result for the combination ($\beta_{12}$, 
$\beta_{13}$) is depicted in \gfig{fig:beta12_vrs_beta13}.
As we can see, the JUNO-TAO (green) sensitivity has no 
correlation and goes from $\sim -0.03$ to $\sim0.03$ for both 
parameters. At the same time, the DUNE-ND (red) sensitivity 
has a positive correlation that drastically reduces the 
sensitivity of both parameters creating a blind spot that can 
reach ($\beta_{12}$, $\beta_{13}$) = $\pm$(0.14, 0.14) when 
the $\delta_D$ is minimized over (dashed-red line). The
$|\beta_{ij}| \sim \mathcal O(0.15)$ in combination with 
DUNE-ND's  average momentum transfer with 
$\ln^2(|Q_p^2|/|Q_d^2)| \approx 10$ is essentially 
outside the linear approximation regime of
\geqn{eq:LinearSolution}.
The presence of extra transition channels, especially for the muon/electron transition into the tau flavor, at FASER$\nu$2 (blue) generates a negative correlation and reduces the impact to the sensitivity when compared to DUNE-ND. However, the correlations are still important and reduces the FASER$\nu$2 capability to constrain the RG running parameters. Notably, the combined 
result (black curve) takes advantage of the different 
correlations among all experiments and greatly improves 
the sensitivity such that we obtain a projected sensitivity of $|\beta_{12}|, |\beta_{13}| \lesssim 0.015$ at 68\% C.L., well within the linear regime.

Besides the degeneracy between $\beta_{12}$ and $\beta_{13}$,
there are other combinations of parameters that benefit from the 
combined analysis. For illustration, in 
\gfig{fig:beta23_vrs_betadelta} we show the expected 
sensitivity of the combination ($\beta_{23}$, 
$\beta_{\delta}$). The JUNO-TAO experiment (dashed-green) is 
insensitive to $\beta_{23}$ as the $P_{ee}$ channel does not 
depend on the parameter as seen in \geqn{eq:Pee}. For 
that reason, we display the JUNO-TAO contour based on the 
$\Delta\chi^2 = 0.05$ cut. Since both DUNE-ND (red) and FASER$\nu$2 
(blue) depend on both parameters we display the contours with 
the 68\% C.L. cut ($\Delta\chi^2 = 2.3$). The shape of the DUNE-ND curve is a combination of the measurement of the survival probability $P_{\mu \mu}$ and the transition probability $P_{\mu e}$. For the two vertical curves, such behavior is determined by $P_{\mu\mu}$ as well as $P_{\mu e}$, which have no correlation between $\beta_{23}$ and $\beta_\delta$ as shown in \geqn{eq:Pmemm}.
The two curved branches are affected by the $P_{\mu e}$ transition probability
whose exact expression is,
\begin{equation}
  P_{\mu e} 
=
  \sin^2 \left(\frac{\Delta \delta}{2}\right)
 \sin^2 2 \theta_{13} 
 \sin^2\theta_{23}^p,
\end{equation}
where $\Delta \delta\equiv \delta(Q^2_d)-\delta(Q^2_p)$ and $\theta_{23}^p\equiv \theta_{23}(Q^2_p)$.
For $\beta_{23}\approx -0.1$, the linear regime is not valid anymore.
Together with $\ln(Q^2_p/Q^2_0)\approx 10$, a large $\beta_{23}$ leads to $\sin^2 (\theta_{23} + 
\Delta \theta_{23}) \approx 0$, which drastically reduces the sensitivity to $\beta_\delta$. At the 
same time, FASER$\nu$2 is also not capable of constraining $\beta_\delta$ inside the linear regime. 
Fortunately, the combined result (black curve) provides a strong constraint on the parameters 
such that $|\beta_{23}| \lesssim 0.01$ and $|\beta_\delta| < 0.045$ at 68\% C. L. 
For completeness, we checked the impact on the sensitivity to the variation of 
$\delta_{\rm D}$, we find that the variation of the Dirac CP-phase has 
negligible effect on the sensitivity curves. For this reason, 
we only show the result with $\delta_D$ prior in
\gfig{fig:beta23_vrs_betadelta}.

\section{Discussion and Conclusions}
\label{sec:conclusions}

New physics related to neutrinos may arise at intermediate 
scales and can generate a RG running of the 
neutrino mixing angles. In the linear regime, running can be 
parametrized by model-independent $\beta_X$ parameters.
Interestingly, a zero-distance transition probability 
appears due to the mismatch of production and detection 
momentum transfer and allows for testing new physics in 
short-baseline detectors.

We show that the presence of several correlation 
among the $\beta_X$ parameters makes it a difficult task for 
any short-baseline experiment alone to model-independently 
constrain the RG running parameters.
However, combining experiments with 
several detection channels and energies such as JUNO-TAO, 
DUNE-ND, and FASER$\nu$2 makes it possible to greatly
reduce such degeneracies and provide constrains on 
$|\beta_{12}|, |\beta_{13}| \lesssim 0.015$, $|\beta_{23}| 
\lesssim 0.01$, and $|\beta_{\delta}| < 0.045$ at 68\% C.L.

\section*{Acknowledgements}
PBD acknowledges support from the US Department of Energy under Grant Contract DE-SC0012704.
SFG is supported by the National Natural Science
Foundation of China (12425506 and 12375101).
SFG is also an affiliate member of Kavli IPMU, University of Tokyo.
This work is also supported by State Key Laboratory of Dark Matter Physics.
CFK is supported
by IBS under the project code IBS-R018-D1. 
PP is supported by Coordenação de Aperfeiçoamento de Pessoal de Nível Superior 
- Brasil (CAPES) - Finance Code 001, with grant number 33003017. PP was also 
supported by FAEPEX, grant number 2404/25 FAEPEX/UNICAMP.

\bibliography{nuMultiRG}
\bibliographystyle{utphysGe}

\end{document}